\RequirePackage{amsmath}
\documentclass[runningheads]{llncs}
\usepackage{graphicx}
\usepackage{multirow,amssymb,mathtools}
\usepackage{booktabs}
\usepackage[caption=false]{subfig}
\begin{document}
\title{Deep Recurrent Modelling of Stationary Bitcoin Price Formation Using the Order Flow}
\titlerunning{Deep Stationary Modelling Using Order Flow}
\author{Ye-Sheen Lim \and Denise Gorse}
\authorrunning{Ye-Sheen Lim \and Denise Gorse}
\institute{University College London - Computer Science \\
Gower Street, London, WC1E 6BT}
\maketitle              % typeset the header of the contribution
\begin{abstract}
In this paper we propose a deep recurrent model based on the order flow for the stationary modelling of the high-frequency directional prices movements. The order flow is the microsecond stream of orders arriving at the exchange, driving the formation of prices seen on the price chart of a stock or currency. To test the stationarity of our proposed model we train our model on data before the 2017 Bitcoin bubble period and test our model during and after the bubble. We show that without any retraining, the proposed model is temporally stable even as Bitcoin trading shifts into an extremely volatile "bubble trouble" period. The significance of the result is shown by benchmarking against existing state-of-the-art models in the literature for modelling price formation using deep learning.
\end{abstract}

\section{Introduction}

The aim of this paper is to investigate the stationary modelling of price formation using deep learning approaches. \emph{Price formation} is an important area in the study of market microstructure, concerning the process by which asset prices form. When modelling price formation in practice, one of the biggest concerns is the \emph{stationarity} of the model. Stationarity is the ability of a model to maintain prediction performance not just out-of-sample, but across a range of periods where the underlying process that generates the data undergoes drastic changes. The financial market is subject to chaotic shift in regimes, and, as a consequence a model that is trained and tested in a particular period is not guaranteed to perform as well if some unobservable underlying process of the financial market causes a drastic shift in the statistical properties of the data. Also, the stationary of price formation modelling is of much interest to financial academia as it is tied closely to the study of the financial markets as a complex dynamical system.

In this paper, we propose a deep recurrent model for modelling the price formation of Bitcoin using the \emph{order flow}. The work is novel in that we are the first to employ an order flow based deep learning model for the modelling of price formation. The order flow is the microsecond timestamped sequence of events arriving at an exchange. Each event is an order placed by a market participant; therefore, order flow is the main endogenous driver of the eventual rise and fall of prices we see on Google Finance charts, for instance. We formulate the \emph{price formation} modelling problem as the forecasting of \emph{high-frequency} \emph{directional price movements}, as is common in quantitative finance literature \cite{bacry2014hawkes,cont2010stochastic}. Bitcoin data is used in our experiment due to the ease of obtaining the high-frequency form of such data as opposed to equivalent data for other financial assets. Also, we are able to obtain Bitcoin data covering periods before and after an extremely volatile bubble period, which is crucial for allowing us to study the stationarity of the models. We train our proposed order flow model on this data and show that our model is able to display the very desirable property of stationarity. We in addition implement state-of-the-art deep learning models from price formation modelling literature, and benchmark our proposed model against them to show the significance of the results.

The paper is organised as follows. In Section 2, we begin by providing a background to the financial concepts touched upon in this paper. In Section 3, we present an overview of related work in the existing literature. Our proposed method and the benchmarks will be presented in Section 4. Sections 5 and 6 cover the data acquisition, experimental setup and results. Finally we conclude with a discussion in Section 7.

% vbox error

\section{Financial Background}

Most, if not all, modern electronic stock or currency exchanges are driven by \emph{limit order books} (LOBs). The electronic LOB is a platform that aggregates the quantities market participants desire to buy or sell at different prices. Most trading activities revolve around the lowest sell and highest buy prices. Readers are directed to \cite{gould2013limit} for a comprehensive introduction to LOBs.

Any exchange that uses a LOB is order-driven, such that any trader can submit \emph{limit orders} (LO) to buy or sell a quantity of an asset at a specific limit price. If the order cannot be satisfied (at the specified limit price or better) on arrival to an exchange, then the LO is added to the LOB to be matched against subsequent orders arriving at the exchange. LOs in the LOB can also be cancelled at any time using a cancellation order (CO). Traders can also submit a \emph{market order} (MO), which has no limit price and is always immediately executed at the best price in the LOB. The sequential stream of order book events is called the \emph{order flow}.

Although there are many useful measures that can be computed from LOBs and order book events, of interest to this paper are the \emph{mid-price}, \emph{best bid}, \emph{best ask} and the \emph{relative price}. The \emph{best bid} and \emph{best ask} are the highest buying and lowest selling prices in the LOB respectively. The \emph{mid-price} is the mean of the best ask and best bid, essentially the mid-point between the highest buying and lowest selling prices in the LOB. The \emph{relative price} is the number of \emph{ticks} between any two prices, where a \emph{tick} is the lowest price increment or decrement allowed by the exchange.

\hfill

\section{Related Work}

Theory-driven modelling of high-frequency (HF) price movements is an extensively researched topic. These approaches usually apply well defined stochastic models, chosen based on empirical analysis and market theories, for modelling HF price movements as a function of different measures of the LOB and order book events. Selected works in this area include \cite{bacry2014hawkes,toke2011introduction,cont2010stochastic}. The advantage of these approaches is their ability to simultaneously produce probabilistic forecasts of high-frequency price movements and confidently explain the predictions using the well-defined theory-driven models. However, major drawbacks include reliance on parametric models, intractability of the models and the lack of generalisation power of the models.  

Data-driven modelling of HF price movements is a relatively recent area of research, especially so in the area of deep learning due to the difficulty of obtaining enough data to train a deep learning model. State-of-the-art models in the literature are based on taking dynamic snapshots of the limit order book to model price movements \cite{sirignano2019universal,tsantekidis2017using}, where a snapshot of the limit order book (LOB) is the price and quantity of a given number of highest bid and lowest ask prices, at a given point in time. It has been shown that these LOB snapshot models can be augmented with market order arrival sequences to improve performance \cite{dixon2018sequence}. Among these existing work, only \cite{dixon2018sequence} performed an analysis on the stationarity of the model. We will later show that while these benchmark models are powerful, our proposed order flow model outperforms them and also exhibits stronger stationarity in the forecasting of Bitcoin during the bubble period.

\section{Methods}

Let us denote the directional price movement at time $T+1$ as $y_{i}\in\{0,1\}$, where $y_i=0$ indicates a downward price movement and $y_i=1$ indicates an upward price movement. In our proposed order flow based approach, we want to model the probability distribution of $y_i$ conditioned on a sequence of irregularly spaced order flow events $\mathbf{x}_i$ of length $T$:

\begin{equation}
	p(y_i | \mathbf{x}_{i,1}, \mathbf{x}_{i,2}, \mathbf{x}_{i,3}, \dots \mathbf{x}_{i,T})
\label{eq:model}
\end{equation}

\noindent
where $\mathbf{x}_{i,t}$ is an order event (e.g. market order, limit order, cancellation order). Each event $\mathbf{x}_{i,t}$ can be described as the tuple $\mathbf{x}_{i,t} =\{x\}^{(j)}_{i,t}$, where:

\begin{itemize}
  \item $x^{(1)}_{i,t} \in \mathbb{N}$ is the number of milliseconds between the arrival of $\mathbf{x}_{i,t-1}$ and $\mathbf{x}_{i,t}$
  \item $x^{(2)}_{i,t} \in \mathbb{N}$ is the hour of the arrival of $\mathbf{x}_{i,t}$ according its timestamp
  \item $x^{(3)}_{i,t} \in \mathbb{R}^+$ is the size of the order $\mathbf{x}_{i,t}$
  \item $x^{(4)}_{i,t} \in \{1,2,3\}$ is the categorical variable for $\mathbf{x}_{i,t}$ being a limit order, market order or cancellation order
  \item $x^{(5)}_{i,t} \in \{1,2\}$ is the categorical variable for $\mathbf{x}_{i,t}$ being a buy or sell order
  \item $x^{(6)}_{i,t} \in \mathbb{N}^+$ is the relative price of the order $x_{i,t}$ divided by the tick (if $x^{(5)}_{i,t} = 1$ then the price is relative to the highest buy price in the LOB, and if $x^{(5)}_{i,t} = 2$ then it is relative to the lowest sell price)
\end{itemize}

We compute the probability distribution of Equation \ref{eq:model} using a softmax function as follows:

\begin{equation}
  P( y_i = j \ | \ \mathbf{h}^L_{i,T}, \mathbf{W}^D_j ) = \frac{e^{ z^D_j( {\mathbf{h}^L_{i,T}}, \mathbf{W}^D_j  ) }}{\sum^{K-1}_{k=0} e^{ z^D_k( {\mathbf{h}^L_{i,T}}, \mathbf{W}^D_k  ) } } \ \ ,
\label{eq:p_y}
\end{equation}

\noindent
where $j \in \{0, 1\}$, $\mathbf{h}^L_{i,T}$ is some learnt $L$-layer deep representation of order flow, and $z^D_k$ is the output layer of a $D$-layer fully-connected neural network. The representation $\mathbf{h}^L_{i,T}$ is the output of a deep $L$-layer recurrent neural network taken at the end of a length $T$ order flow:

\begin{equation}
  \mathbf{h}^l_{i,T} =
  \begin{cases}
    h( \mathbf{h}^{l-1}_{i,T}, \mathbf{h}^{l}_{i,T-1}, \Theta^l ) & \text{if $1 < l \leq L$} \\
    h( \mathbf{x}_{i,T}, \mathbf{h}^{l}_{i,T-1}, \Theta^l )  & \text{if $l=1$} \\
  \end{cases}  \ \ ,
\label{eq:h_l}
\end{equation}

\noindent
where $h(.)$ is a function implementing a recurrent neural network with long short-term memory (LSTM) cells, $\Theta^l$ are LSTM parameters to be fitted, and $\mathbf{x}_{i,T}$ will be soon addressed. Since LSTM cells are commonly implemented RNN components in the literature, we will not discuss their architecture here and direct readers instead to \cite{hochreiter1997long}. 

For each order $x_{i,t}$ all non-ordinal categorical covariates are embedded into a multidimensional continuous space before feeding into the inputs of the RNNs. The embeddings can be described as follows:

\begin{equation}
 e(q_{i,t}) = g\left( {\mathbf{U}_q}^\intercal o(q_{i,t}) +\mathbf{b}_q \right)
\label{eq:emb}
\end{equation}

\noindent
where $o(.)$ is a function mapping the categorical features to one-hot vectors, $g(.)$ is some non-linear activation function, and $\mathbf{U}_q$ and $\mathbf{b}_q$ are parameters to be fitted.

Then, the parameters of the model $\mathbf{W}^d_k$ and $\mathbf{\Theta}^l$, as well as the weights and bias terms in the embedding layers for all covariates, can be fitted using any variant of the stochastic gradient descent optimisation algorithms by minimising the negative log-likelihood:

\begin{equation}
  \mathcal{L}(\mathbf{y}) = - \frac{1}{N} \sum^N_{i=1} \sum^{K-1}_{j=0} \mathbb{I}_{y_i=j} \ log \ p(j)
\label{eq:ce}
\end{equation}

\noindent
where $\mathbb{I}$ is the identity function that is equal to one if $y_i=j$ and is zero otherwise, $K$ is the number of classes, and $N$ is the size of our dataset.

The performance of the model will be evaluated using the Matthews correlation coefficient (MCC) \cite{powers2011evaluation}. We choose this metric as it has a very intuitive interpretation, it handles imbalanced classes naturally. For binary classification, the metric lies in the range $(-1,1)$ where $1$ indicates a perfect classifier, $-1$ indicates a completely wrong classifier and $0$ means the classifier is doing no better than making random predictions (making this measure very useful in the context of quantitative trading). As it summarises the confusion matrix into one balanced and intuitively interpretable measure, it allows us to perform concise and extensive comparisons without needing to delve into the relative contributions of different elements of the confusion matrix.

We benchmark the performance of our order flow model against two state-of-the-art models found in the literature applying deep learning to high-frequency price modelling. These models can be described as follows:

\begin{itemize}
 \item \textbf{Benchmark 1} \cite{dixon2018sequence} models the probability distribution of $y_i$ as in Equation \ref{eq:model}, but each element $\mathbf{x}_{i,t}$ in the irregularly spaced sequence $\mathbf{x}_i$ is defined as follows:
  \begin{equation}
    x_{i,t} = ( b_{i,t}^1, b_{i,t}^2, \dots b_{i,t}^S, s_{i,t}^1, s_{i,t}^2, \dots s_{i,t}^S, \alpha_{i,t}^b, \alpha_{i,t}^s)  \ \ ,
  \end{equation}
  where $b^j_{i, t} = (\pi^j_{i,t}, \omega^j_{i,t}) $ is the tuple of price $\pi$ and volume $\omega$ at the $j$'th highest buy price, $s^j_{i, t} = (\phi^j_{i,t}, \kappa^j_{i,t}) $ is the tuple of price $\phi$ and volume $\kappa$ at the $j$'th lowest sell price, and $S$ is the number of highest buy or lowest sell prices we are considering in the snapshot. The market order (MO) rate on the buy side of the LOB, $\alpha_{i,t}^b$, is computed by dividing the number of buy MOs in the period between $t=0$ and $T$ by the total number of orders that make up the volume in the highest bid price. The MO rate on the sell side of the LOB, $\alpha_{i,t}^s$, is computed similarly.
  \item \textbf{Benchmark 2} \cite{sirignano2019universal,tsantekidis2017using} models the probability distribution of $y_i$ as in Equation \ref{eq:model}, but each element $\mathbf{x}_{i,t}$ in the irregularly spaced sequence $\mathbf{x}_i$ is here defined similarly to Benchmark 1 but without the MO rates:
  \begin{equation}
    x_{i, t} = ( b_{i,t}^1, b_{i,t}^2, \dots b_{i,t}^S, s_{i,t}^1, s_{i,t}^2, \dots s_{i,t}^S )  \ \ ,
  \end{equation}
\end{itemize}

\noindent
For both these benchmark models, the probability distributions of $y_i$ are computed using softmax functions that take as inputs deep representations of $\mathbf{x}_i$ learnt from recurrent neural networks.

\section{Experimental Set-Up}

The dataset for the experiments was obtained from Coinbase, a digital currency exchange. Through a websocket feed provided by the exchange, we log the real-time message stream containing trades and orders data in the form of order flow for BTC-USD, BTC-EUR and BTC-GBP from 4 Nov 2017 to 29 Jan 2018. Since these are raw JSON messages, the dataset for training the model cannot be obtained directly from the messages. To build the datasets, we had to reconstuct the limit order book by sequentially processing the JSON messages.

During the order book reconstruction, we build the dataset in real-time using the following method. Before we begin building the dataset, we have to "warm up" the order book using messages from 4 Nov 2017 - 5 Nov 2017. This needs to be done because we are starting from an empty book that needs to be sufficiently populated for us to extract sensible data. Now we begin tracking the mid-price. If the mid-price moves after an order $x_{i,T+1}$ arrives, then the mid-price is stored as the target variable class $y_i \in \{0,1\}$, for downward or upward mid-price movements respectively. We ignore any events that do not move the price on arrival since we are only interested in modelling price formation (predicting up or down price movements). After a mid-price change is registered, we look back into the order flow and the limit order book to obtain the measures needed to build the datasets for our proposed model and for the benchmark models.

Each dataset is then split into training, validation and test sets by dates to avoid any look-ahead bias. About 1.1 million datapoints between 6 Nov 2017 and 16 Nov 2017 are taken as the training set. Cross-validation and early-stopping checks are performed on a set taken from between 17 Nov 2017 to 20 Nov 2017 containing about 0.55 million datapoints. The rest is kept strictly for testing and is never seen by the model aside from the final testing phase, giving us about 7.3 million test points in total.

We set up the order flow and benchmark models as described in Section 4. Then we fit the parameters of individual models using the Adam optimisation algorithm with dropout on all non-recurrent connections. Cross-validation is performed with Bayesian hyperparameter tuning to find the number of recurrent layers, the LSTM state size, the number of dense layer in the output, the size of each output dense layers, and the embedding dimensions. Embedding is not used for the benchmark models since they do not contain categorical variables. We then make predictions on the test sets. The results are then grouped by date and we compute the Matthews correlation coefficient (MCC) for each date.

\section{Results}

\subsection{Comparison of Model Performances}

\begin{figure}[!ht]
     \subfloat[BTC-USD\label{fig:usd}]{%
       \includegraphics[width=0.49\textwidth]{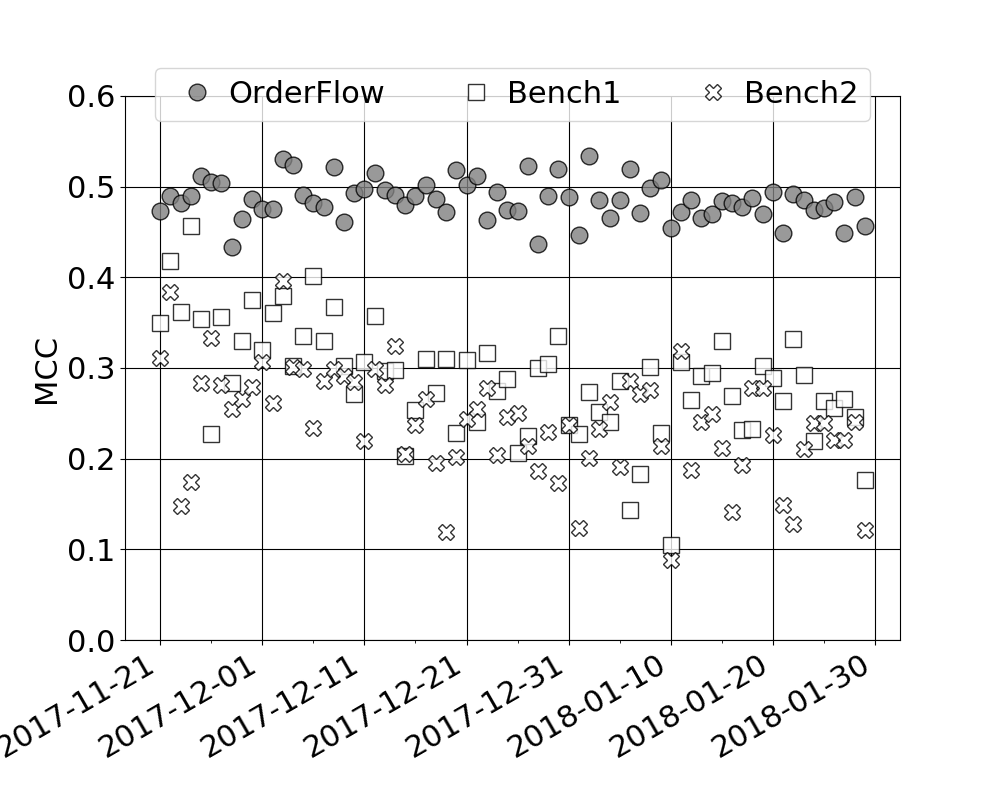}
     }
     \hfill
     \subfloat[BTC-EUR\label{fig:eur}]{%
       \includegraphics[width=0.49\textwidth]{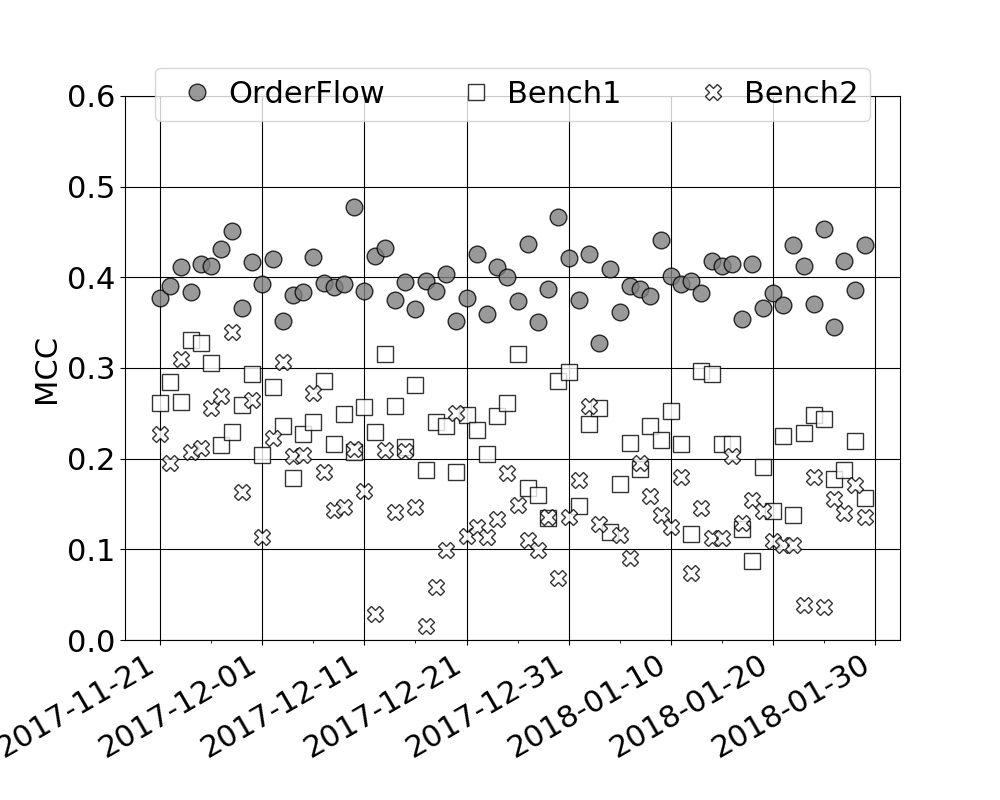}
     }
     \hfill
     \centering
     \subfloat[BTC-GBP\label{fig:gbp}]{%
       \includegraphics[width=0.49\textwidth]{figs/eur}
     }
     \caption{MCC plots for models trained and tested on BTC-USD, BTC-EUR and BTC-GBP respectively}
     \label{fig:usd_eur_gbp}
\end{figure}

Let us evaluate the relative performance of the various models, with results presented in Figures 1a-c. Figure \ref{fig:usd} shows the MCC over each day for models trained and tested on the BTC-USD dataset. Note that the abbreviation \emph{OrderFlow} refers to our proposed model, while \emph{Bench1} and \emph{Bench2} refer to Benchmark 1 and Benchmark 2 respectively. We first note in addition that in all the figures of Figure \ref{fig:usd_eur_gbp} we can see that \emph{Bench1} (which augments the model with market order rates) outperforms \emph{Bench2} (which only models the price movements on the LOB snapshot) throughout the whole test period. However, we observe that throughout the test period, the order flow model outperforms those of both benchmark models. Although not presented, we also performed paired Student t-tests on the null hypothesis that there is no difference between the test results of the order flow model and that of benchmark models individually. In each test, the null hypothesis is rejected with very high confidence intervals. Similarly, Figure \ref{fig:eur} and Figure \ref{fig:gbp} show the test performance for models trained and tested on BTC-EUR and BTC-GBP, respectively. We are again able to visually verify that models trained on order flow outperform the benchmark models, and once again paired Student t-tests reject with high confidence the null hypotheses that \emph{OrderFlow} test results are no different from the benchmark models \emph{Bench1} and \emph{Bench2}.

\subsection{Analysis of Model Stationarity}

\begin{figure}[ht!]
    \centering
    \includegraphics[width=0.65\columnwidth]{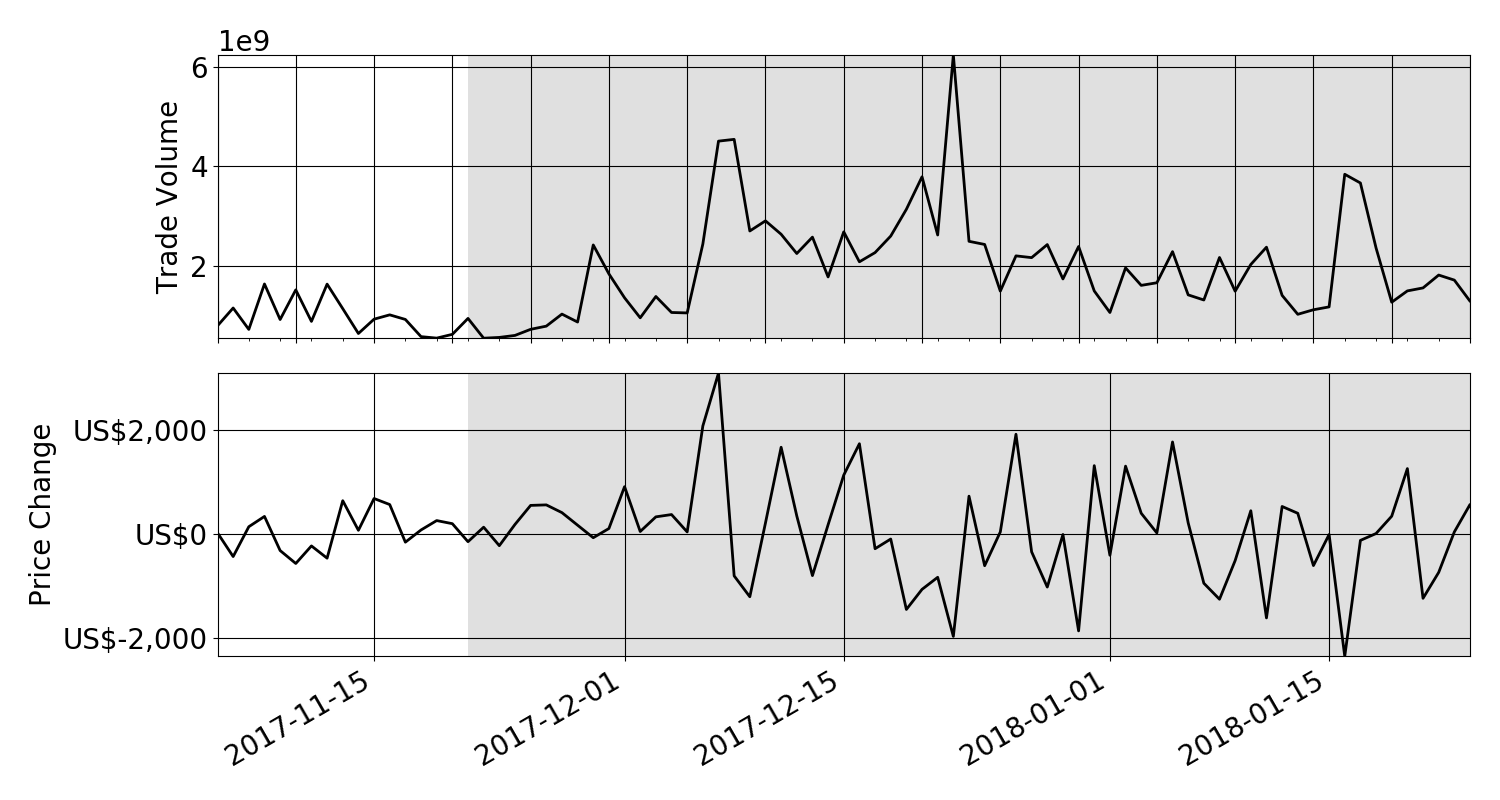}
    \caption{Plot of: i) volume of trading activity (Trade Volume), ii) BTC-USD 1-day lagged difference prices (Price Change), the shaded area being our test period 21 Nov 2017 - 29 Jan 2018}
    \label{fig:btc_trading}
\end{figure}

The range of dates in our test period covers the climax of the Bitcoin bubble where the price of Bitcoin rapidly peaks to an all-time-high and subsequently bursts \cite{kreuser2018bitcoin}. We can see in Figure \ref{fig:btc_trading} that compared to the training and validation period, our test period for BTC-USD (and in fact for BTC-EUR and BTC-GBP also) corresponds to a shift in regime characterised by more volatile price changes and much higher trading activity. With this test period, we can evaluate the stationarity of our proposed order flow model and the benchmark model.

From Figures \ref{fig:usd} and \ref{fig:eur}, we can in all cases (BTC-USD, BTC-EUR, BTC-GBP) visually observe that while the performance of \emph{Bench2} degrades less quickly than \emph{Bench1}, the model trained on order flow is the only one to display substantial stationarity. This implies that the representation learnt from the order flow is transferable from a non-volatile to an extremely volatile period, suggesting that the learnt representation encodes some sort of temporally universal information about Bitcoin price movements. On the other hand, we can visually observe that the benchmark models struggle to maintain performance in the volatile period. To scientifically evaluate and statistically compare the stationarity of the order flow model and the benchmark models, we fit a linear regression model on the test results of each model. This will then allow us to compare the rate at which the performance of each model degrades over time.

Specifically, we fit a simple linear regression on the MCC over the test period dates. Table \ref{table:slopes} shows the corresponding slope coefficients and p-values for each model trained and tested on a particular currency pair. Although the model trained on order flow has negative slope coefficients for all datasets, implying some degradation in performance over time, the p-values reveal that the coefficient is not statistically significant. However, for the benchmark models we can very confidently reject the null hypotheses that the slopes are zero, meaning that we can use the negative coefficients as strong evidence for performance degradation over time.

\begin{table}[ht!]
\centering
\begin{tabular}{||c | c | c c||}
 \hline
 Dataset & Model & Coeff & p-value \\
 \hline\hline
 \multirow{5}{*}{BTC-USD} & Order Flow Model & $-2.41e^{-4}$ & $5.87e^{-2}$ \\
 & Benchmark 1 & $-2.76e^{-3}$ & $1.27e^{-13}$  \\
 & Benchmark 2 & $-1.63e^{-3}$ & $8.27e^{-4}$ \\
 \hline
  \multirow{5}{*}{BTC-EUR} & Order Flow Model & $-5.67e^{-5}$ & $7.50e^{-1}$ \\
 & Benchmark 1 & $-2.02e^{-3}$ & $1.27e^{-8}$  \\
 & Benchmark 2 & $-2.01e^{-3}$ & $1.19e^{-5}$ \\
 \hline
  \multirow{5}{*}{BTC-GBP} & Order Flow Model & $-4.01e^{-4}$ & $1.68e^{-1}$ \\
 & Benchmark 1 & $-2.36e^{-3}$ & $1.29e^{-11}$  \\
 & Benchmark 2 & $-1.35e^{-3}$ & $4.22e^{-5}$ \\
 \hline
\end{tabular}
\caption{The table shows the slope coefficients and p-value of MCC regressed on dates in the test period for individual models that are: i) trained and tested on BTC-USD, ii) trained and tested on BTC-EUR, iii) trained and tested on BTC-GBP.}
\label{table:slopes}
\end{table}

\section{Analysis of Model Universality}

Although we set out primarily to study stationarity in our model, it is in addition possible to show that the representations learnt from the order flow exhibit a hint of the very valuable property of universality \cite{sirignano2019universal}, the ability to learn market structures which to some degree generalise across asset classes.

Table \ref{tab:universal} shows the drop in performance on the out of sample test set, when training on one currency pair and testing on the others, is considerably less when using the order flow model than the benchmark models, demonstrating the above-mentioned hint of universality. This innate ability to generalise is most evident when training on BTC-USD and least when training on BTC-GBP. This is likely due to the different volumes traded: the trading volumes of BTC-USD, BTC-EUR, and BTC-GBP between the start of the training period (6 Nov 2017) and the end of the test period (29 Jan 2018) are, respectively, $151.6e9$, $20.7e9$, and $1.5e9$  -- when an asset is heavily traded there are more activities at the order book level, resulting in a richer order flow and hence a richer dataset that helps to avoid overfitting.

\begin{table}[ht!]
\centering
\begin{tabular}{||l|cc|cc|cc||}
\hline
Trained & \multicolumn{2}{c|}{BTC-USD} & \multicolumn{2}{c|}{BTC-EUR} & \multicolumn{2}{c|}{BTC-GBP} \\ \hline
Tested  & BTC-EUR       & BTC-GBP      & BTC-USD       & BTC-GBP      & BTC-USD       & BTC-EUR      \\ \hline
Order Flow & 9.299  & 20.306  & 19.606 & 17.387  & 28.816 & 32.232  \\
Benchmark 1  & 59.857 & 71.749  & 74.368 & 80.000  &  67.157 & 67.642  \\
Benchmark 2 & 91.017 & 81.869  & 99.619 & 84.049  &  80.432 & 70.483  \\ \hline
\end{tabular}
\caption{For each model (leftmost column), the presentation tabulates the mean percentage drop $(\%)$ in test MCC between training and testing on the same currency pair, and training on a given currency pair and testing on the other currency pairs.}
\label{tab:universal}
\end{table}

\section{Discussion}

We have presented a model for the prediction of the directional price movements of Bitcoin using the order flow (which is the raw market data). We showed that the model is able to partially achieve a temporally universal representation (the very valuable property of stationarity) of the price formation process such that even when the statistical behaviour of the market changes dramatically (here, after the bursting of the Bitcoin bubble), the model remains relatively unaffected. We also show that the stationarity performance of our proposed order flow model is substantially better than benchmark deep models obtained from the existing literature. A secondary analysis of the results also hints at a universality property of our proposed model, and this too is benchmarked against the same deep models from the existing literature, to the benefit of our order flow model.

For future work, since our predictions give encouragingly high MCC values, it will be of interest to apply black-box feature explainers such as Shapley Additive Explanations \cite{lundberg2017unified} to address the interpretation issue of these data-driven models and understand exactly what it is that drives price formation across BTC-USD, BTC-EUR and BTC-GBP, and (in future work) other cryptocurrency pairs. This would provide a data-driven view of the market microstructural behavior of cryptocurrencies. Also of interest is what we can learn about the cryptocurrency market microstructure from analysis of the embeddings of categorical features of the order flow.

%Finally, we might ask to what degree the current models could be useful as pre-trained networks for use in related problems. In our analysis of universality, we show that it is possible to use the order flow model for the prediction of directional price movements for cryptocurrency pairs other than those on which the models were trained. However, we might go further and ask if the learnt representations, with their displayed degree of stationary, could be used to initialise training, and by this route lead to faster convergence, for directional forecasting problems outside of the cryptocurrency context. Perhaps most interestingly, we could investigate whether models trained on high frequency data have learnt anything sufficiently general about the market microstructure so as to be relevant to a different (lower) timescale, where less data would be available and where a pre-trained learner would potentially be very valuable.

\bibliographystyle{splncs04}
\bibliography{ref}

\end{document}